\input harvmac
\input epsf

\Title{\vbox{\hbox{PUPT-???}
\hbox{\tt hep-th/9512225}}}
{The Master Field in Generalised $QCD_2$}
\bigskip
\centerline{Rajesh Gopakumar\foot{E-mail: rgk@puhep1.princeton.edu}}
\bigskip\centerline{\it Joseph Henry Laboratories,}
\centerline{\it Princeton University,}
\centerline{\it Princeton, New Jersey 08544.}
\def\Tr{\rm Tr \, }

\def\1{$1/N$\,}

\def\ad{a^{\dagger}}

\def\vev #1{\left\langle \Omega| #1 |\Omega\right\rangle}

\def\figin{\epsfcheck\figin}\def\figins{\epsfcheck\figins}
\def\epsfcheck{\ifx\epsfbox\UnDeFiNeD
\message{(NO epsf.tex, FIGURES WILL BE IGNORED)}
\gdef\figin##1{\vskip2in}\gdef\figins##1{\hskip.5in}
instead
\else\message{(FIGURES WILL BE INCLUDED)}%
\gdef\figin##1{##1}\gdef\figins##1{##1}\fi}
\def\DefWarn#1{}
\def\figinsert{\goodbreak\midinsert}
\def\ifig#1#2#3{\DefWarn#1\xdef#1{fig.~\the\figno}
\writedef{#1\leftbracket fig.\noexpand~\the\figno}%
\figinsert\figin{\centerline{#3}}\medskip\centerline{\vbox{\baselineskip12pt
\advance\hsize by -1truein\noindent\footnotefont{\bf Fig.~\the\figno:}
#2}}
\bigskip\endinsert\global\advance\figno by1}

\vskip .5in
As an illustration of the formalism of the master field we consider
generalised $QCD_2$. We show how Wilson Loop averages for an arbitrary 
contour can be computed explicitly and with some ease. A generalised 
Hopf equation is shown to govern the behaviour of the eigenvalue 
density of Wilson loops. The collective field description of the theory
is therefore deduced. Finally, the non-trivial master gauge field
and field strengths are obtained. These results do not seem easily
accessible with conventional means.

\Date{\it December 1995}

\newsec{Introduction}
We are faced with a paucity of techniques for understanding, 
quantitatively, the dynamics of strongly interacting field 
theories. One of the most intriguing, and probably the least 
well understood, proposals has been to study the theory in the 
so called large $N$ limit. Here N refers to the size of the gauge group 
which may be, say, SU(N) or SO(N). It is well known that these theories 
do simplify in the large $N$ limit. The free energy and correlation functions 
have a well defined expansion in powers of $1\over{N}$. Moreover the 
leading term --- the planar limit --- is thought to capture the 
essential physics. 

One of the intriguing features of this limit is the existence of a
master field \ref\witten{ E.  Witten, {\it in Recent Developments in 
Gauge Theories}
eds. G. 'tHooft et. al. Plenum Press, New York and London  (1980).}. 
This notion arises from the observation that the large $N$  limit
is in some sense a classical limit with $1\over{N}$ playing the role of 
$\hbar$. The factorisation of products of gauge invariant observables,  
with corrections of $O(\1)$ bears this out \ref\mig1{A.  Migdal,  
{\it Ann. Phys.} {\bf 109},365 (1977).}. The master field 
refers to this \lq \lq classical'' configuration that dominates the 
path integral. Its knowledge would enable one to compute gauge invariant 
quantities without performing the functional integral. We simply evaluate
them at this point in field space. In fact, it can be argued that by a 
suitable gauge transformation the master gauge field
$\bar A_\mu(x)$ can be made space time independent. Thus, for example,
in $QCD_4$ we need  obtain just four
$\lq \lq \infty\times \infty "$ matrices  $\bar A_\mu$!

Tantalising as this prospect may seem, it has proved rather elusive to
work with. We need to understand what these $\lq \lq \infty\times \infty 
"$ matrices really are. And then again we need effective ways to 
compute them. 

Recently, it has been possible to get a better hold on this concept
\ref\us{R.Gopakumar, D.J.Gross,{\it Nucl. Phys.} {\bf B451}, 379 (1995)},
\ref\mrd1{M.R.Douglas hep-th/9409098, {\it Phys. Lett.} 
{\bf 344B}, 117, (1995)}
\ref\vol{I.Ya.Aref'eva,I.V.Volovich hep-th/9510210}. 
The mathematical formalism of non-commutative
probability theory \ref\voic{D. V. Voiculescu, K.J. Dykema and 
A. Nica {\it Free Random Variables} AMS , Providence (1992)} 
seems to be the right setting for the 
master field. Let us briefly summarise the relevant facts. 

The master field can be thought of as an operator living in a Fock space
generated by creation operators obeying no relations. 
In other words, a Fock space spanned by the states
\eqn\fock{(\hat \ad_{i_1} )^{n_{i_1}}(\hat \ad_{i_2} )^{n_{i_2}} \dots (\hat
\ad_{i_k})^{n_{i_k}}|\Omega \rangle }
where
\eqn\boltz{\hat  a_i |\Omega \rangle = 0, \ \  \hat a_i \hat \ad_j = \delta_{ij}.}
Here the subscripts $i$ can take either discrete or continuous values.
In an $n$-matrix model $i$ will run from $1$ to $n$. A general operator,
the master field included, will be built out of these $\hat a_i$'s and
$\hat \ad_j$'s. 

In the theory of non-commutative probability, a very special place is 
occupied by the so called free random variables. They are akin to 
independent random variables in usual probability theory. They are
best thought of as the large $N$ limit of matrices with independent 
distributions. Thus for an independent $n$-matrix model the master fields
$\hat M_i$ corresponding to the matrices $M_i$ are free random variables.
They turn out to have a simple realisation on this fock space.
\eqn\mi{\hat M_i=\hat a_i + \sum_{n=0}^\infty M_{n+1}^{(i)}\hat a_i^{\dagger n}.}
An arbitrary invariant correlation function of the theory is then computable
using
\eqn\corr{\langle \Tr [M^{n_1}_{i_1} M^{n_2}_{i_2}\dots M^{n_k}_{i_k}]\rangle=\vev{\hat M^{n_1}_{i_1} \hat M^{n_2}_{i_2}\dots \hat M^{n_k}_{i_k}}.}
It turns out that the $M_n^{(i)}$ have a rather simple physical interpretation.
They are determined solely by the distribution for $M_i$ and are the connected $n$ point Green's functions. In other words, dropping the subscript $i$,
\eqn\conn{M_n = \langle \tr[M^n]\rangle_{\rm conn.}.}
Or equivalently \us \ the generating function for connected Green's functions
\eqn\Mz{zM(z)=1 +\sum_{n=1} M_n z^n}
is the inverse function of the resolvent
\eqn\inv{R(\zeta) = \sum _{n=0}^\infty\zeta^{-n-1}\vev{M^n}.}
A generalisation of eqn.\mi \ to an arbitrary matrix model also exists.
We refer the reader to \us \ for details on this as well as 
equations of motion, examples and more.

However, it has been difficult to compute the master field in this formalism 
other than in cases where the large $N$ correlation functions have 
already been 
obtained by conventional methods such as the saddle point  or
the Schwinger-Dyson equations. It might then appear that our framework merely
allows us to rewrite results of the large $N$  theory in a compact way. 
This is 
not true. As we shall see in this paper the example of generalised $QCD_2$
is an instance where we can profitably apply ideas from the theory of free
random variables to explicitly compute many quantities of interest. Due
to the nonlinearity of the theory, the usual methods, mentioned earlier,
do not appear very tractable. The main notion that is exploited here is 
that of a multiplicative free family of random variables \voic ,\us
--- a notion that we will elaborate on below. It allows us to construct the 
Wilson loop average for an arbitrary contour from knowledge of certain
infinitesimal ones. We also obtain a realisation of the 
generalised Hopf equation as a collective field equation for the theory.
This will allow us to deduce the collective field theory in the general
case of the cylinder.
Finally, one can construct the master 
field strength and gauge field.

Generalised $QCD_2$ \ref\DLS{M.R.Douglas, K.Li, M.Staudacher
{\it Nucl. Phys. } {\bf B420} 118, (1994)},
\ref\GSY{O.Ganor, J.Sonnenschein, S.Yankielowicz,
{\it Nucl. Phys. }{\bf B434} 139, (1995)} 
has been argued to 
be a theory of interest in the search for potential string theory descriptions
of higher dimensional $QCD$. It generalises the heat kernel action of
$QCD_2$. Alternatively, it can be defined as the
general perturbation to topological $YM_2$:
\eqn\gYM{{\cal Z} =\int[{\cal D}A][{\cal D}B]\exp(-\int(\tr(iBF)-\Phi(B))).}
While it doesn't have propagating modes, it is  
nevertheless more complex in behaviour than ordinary $QCD$. For instance,
it exhibits the large $N$ phase transition \ref\DK{M.R.Douglas,V.A.Kazakov
{\it  Phys. Lett.} {\bf 319B} 219, (1993)} on Riemann surfaces of 
almost all 
genera \ref\RY{B.Rusakov, S.Yankielowicz
{\it  Phys. Lett.} {\bf 339B}, 258,(1994)}. 
It therefore seems worthwhile to have at hand,
explicit results analogous to \ref\KK{V. Kazakov and I. Kostov
{\it Nucl. Phys.} {\bf B176},199 (1980)}. 
From this point of view the 
generalised Hopf equation that we obtain below would have bearing on 
a world sheet lagrangian description.
In Section 2 we outline the 
notion of multiplicatively free families and its appearance in 
$QCD_2$ where it enables us to compute arbitrary loop averages.
We proceed in Section 3  to calculate the loop averages $<\tr [U^n]>$ for  
a simple loop which is then sufficient to obtain a general one.
In Section 4 we show that the behaviour of eigenvalues of loops are governed
by the generalised Hopf equation. The collective field theory is then
reconstructed.
Section 5 is devoted to the construction
of the master gauge field  and field strength. Discussions and conclusions
comprise section 6. An appendix bears the burden of a necessary
combinatorial calculation.

\newsec{Loops as Free Random Variables }

We indicate how a knowledge of infinitesimal loops in generalised $QCD_2$
(ordinary $QCD_2$ is a particular case) can be used
to compute arbitrary loop averages. The basic loop entities that we will be
working with are the simple loops. (By a simple loop we shall henceforth mean
a non-self intersecting contour.) Their importance lies in the fact that
a set of simple loops, based at one point and non-overlapping, correspond to  
a family of free random variables ( Fig. 1). 
\ifig\simp{Three Simple Loops}
{\epsfxsize 2.0in\epsfbox{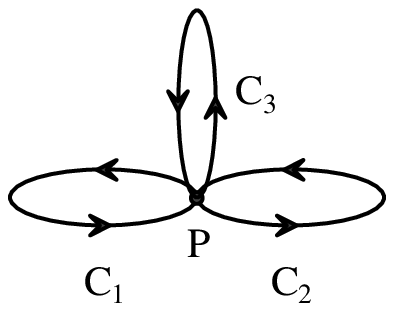}}

To see this, notice that we could work with the heat kernel action
for generalised $QCD_2$ \GSY ,
\eqn\heatk{
\eqalign{& {\cal Z} = \int \prod_L {\cal D}U_L\prod_{\rm plaquettes}
\!\!\!\!Z_P[U_P];\cr 
& Z_P[U_P]= \sum_R d_R\chi_R[U_P] e^{-\Lambda(R)A_P},}}
where $\Lambda(R)$ is an arbitrary function of the Casimirs.
Now the self similar nature of the heat kernel always allows us to 
triangulate the plane such that the contours $C_i$ are the borders
of some of the triangles. Let us denote the $U(N)$ holonomies 
along $C_i$ by $U_i$. Then in any computation of averages of products of 
$U_i$, we can integrate out (only on the plane) all other 
link variables. We are thus left with an equivalent measure ${\cal Z} =\int
\prod_i {\cal D}U_i \prod_i Z[U_i]$, where the product runs over all  the
simple loops, $U_i$. In other words, $U_i$ have independent distributions.

Actually, the $U_i$'s are not just free random variables. They comprise
what is known as a multiplicative free family \voic.  Briefly what 
this means is the following: The product of two free random variables 
with distributions, $\mu_1$ and $\mu_2$
is also a free random variable with some distribution $\mu_3$ denoted by
$\mu_1 \otimes \mu_2$. A one parameter family
of free random variables,  such that $\mu_{t_1} \otimes
\mu_{t_2}=\mu_{t_1t_2}$,
will be called a multiplicative free family. (Or equivalently
$\mu_{s_1} \otimes \mu_{s_2}=\mu_{s_1+s_2}$, if we redefine
the parameter $t \rightarrow s=\log t$.)

Here the area plays the role of the parameter $s$. In other words, for the   
two simple loops $C_1$ and $C_2$ (\simp), 
$\hat{U}_{C_1}(A_1)\hat{U}_{C_2}(A_2)$ 
has the same distribution as $\hat{U}_{C_1\circ C_2}(A_1+A_2)$. This follows
from the self reproducing nature of the heat kernel action \heatk \ together
with its exponential dependence on the area. A more direct argument is made
in \us. 

This allows us to obtain the distribution for a simple loop of finite area by
starting from one of infinitesimal area. The precise way to do this is to
use the S-Transform . This is analogous to the Mellin transform of
ordinary probabilty theory, i.e. it is multiplicative for the product 
of distributions. Therefore for our multiplicative free family, 
$S_{A_1}(z)S_{A_2}(z)=S_{A_1+A_2}(z)$. It then follows that $S_A(z)$
is exponential in $A$.

The function $S(z)$ for a non-commutative random variable $U$
is constructed as follows \voic:
If
\eqn\phij{\phi(j)=\sum_{n=1}^{\infty}\vev{\hat U^n} j^n ,}
then construct the inverse function  $\chi(z)$, i.e. $\phi(\chi(z))=z$.
The S-transform is defined as:
\eqn\Sdef{S(z)={1+z\over z}\chi(z) .}
In the next section we shall compute $S_A(z)$ for infinitesimal $A$ 
and then exponentiate it to obtain that for finite $A$. Then we shall
take the \lq inverse transform' to finally arrive at $<\tr [U^n]>$.

We conclude the section with a sketch of how arbitrary loop averages are 
computed. This requires two ingredients. Firstly, a geometrical 
decomposition of an arbitrary loop into a \lq word' built out of simple,
non-overlapping loops, based at one point. And secondly, that 
the latter are free random variables. 

The first was illustrated at length in \us. Here we shall merely
give an instance of how it works. 
\ifig\nasty{A Loop and its Decomposition into Simple Loops}
{\epsfxsize
4.0in\epsfbox{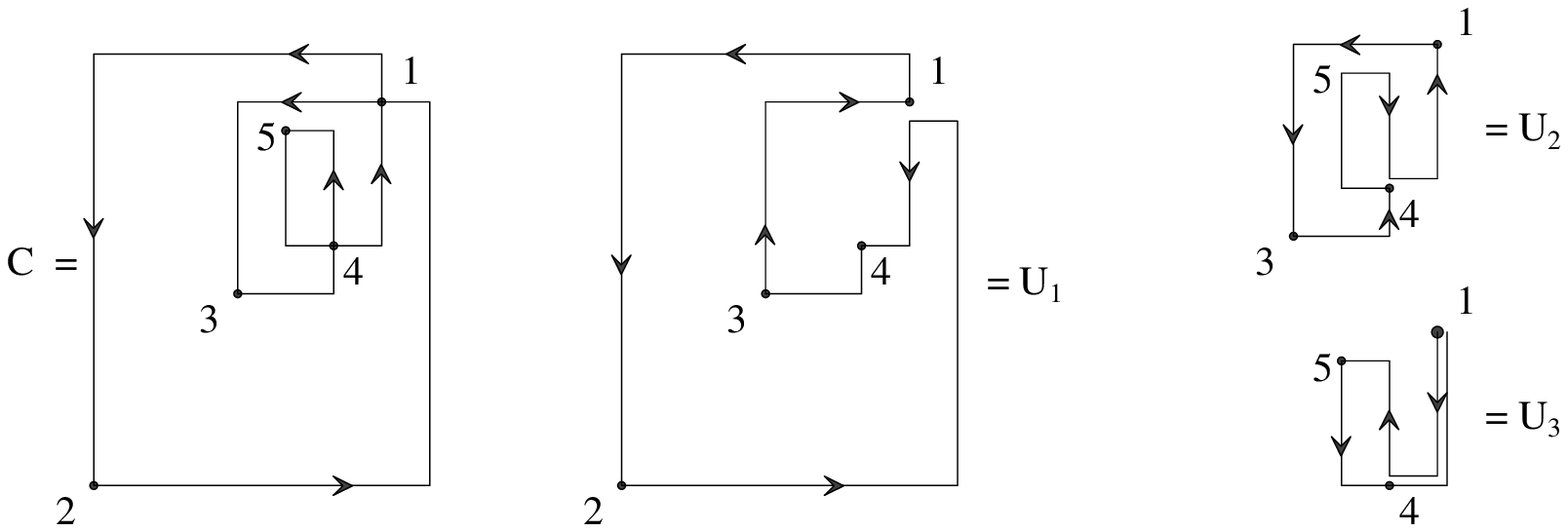}}
With the labelling in \nasty \ we can write the contour as 
\eqn\nasty{\eqalign{
C=&(12)(\overline{21})(13)(34)(45)(\overline{54})(41)\cr
=&\underbrace{(12)(\overline{21})(14)(43)(31)}_{U_1}(13)(34)(41)(13)(34)(45)
(\overline{54})(41)\cr
=&\underbrace{(12)(\overline{21})(14)(43)(31)}_{U_1}\underbrace{
(13)(34)(\overline{45})(54)(41)
}_{U_2}\underbrace{(14)(45)(\overline{54})(41)}_{U_3}\cr
&\underbrace{(13)(34)(\overline{45})(54)(41)}_{U_2} 
\underbrace{(14)(45)(\overline{54})(41)}_{U_3}\underbrace{(14)(45)(\overline{54})
(41)}_{U_3}\cr
= &U_1U_2U_3U_2U^2_3. \cr}}
The main strategy is to introduce backtracking or \lq thin' loops
to peel off the simple loops $U_i$ corresponding to the various windows.

Having decomposed the loop into a word, we can exploit the fact that
the $U_i$ are free. The loop average for \nasty, for instance, reads as
\eqn\nastyavg{\eqalign{
W(C)=&<\tr [U_1U_2U_3U_2U^2_3]>=<\tr U_1><\tr [U_2U_3U_2U^2_3]>\cr
=&<\tr U_1>(<\tr U_2>^2<\tr U^3_3>-<\tr U_2>^2<\tr U_3><\tr U^2_3>\cr
&+<\tr U^2_2><\tr  U_3><\tr U^2_3>).\cr}}
Here we have used a recursion relation for free random variables
(see \us ) to factorise the
word into \lq letters' of the form $<\tr [U^n]>$. Once we evaluate the latter,
we are done.

\newsec{$<\tr [U^n]>$ in Generalised $QCD_2$}

As mentioned earlier, we utilise the S-Transform \Sdef ,\phij. This requires
us to evaluate $<\tr [U^n]>$ for an infinitesimal loop. Again it is the heat
kernel definition that is useful. By definition
\eqn\Wn{W_n \equiv \sum_R d_R\int {\cal D}U \chi_R(U) \tr [U^n]  
e^{-{A\over2}\Lambda(R)} .}
To evaluate the $U$ integral we can use the fermion representation
of U(N) \ref\mrd2{M.R.Douglas, Cargese '93, hep-th/9311130}. 
The characters are now the Slater fermion wavefunctions.
In a second quantised language $\tr [U^n]$ becomes a fermion bilinear and
\eqn\Wferm{\int {\cal D}U\chi_R(U) \tr [U^n]= \langle O|\sum_m 
B^{\dagger}_{n-m}B_m |R \rangle.}
Here $|O\rangle$ is the ground state of the N fermions and {\it not} the 
vacuum.
In terms of Young tableaux this is the state with no boxes. $|R \rangle$
is the fermionic state labelled by integers $\{n_i\}$. These integers are 
related to the number of boxes $h_i$ in the $i$th row of the young tableau
by $n_i=(N-1)/2+1-i+h_i$.  

It is now easy to see that the only representations that contribute in
\Wferm \ have 
\eqn\YT{\eqalign
{h_1=&n-m\cr 
 h_i=& 1 \ \ \  (2 \leq i \leq m+1) \cr}}
with $m$ any integer between $0$ and $n-1$. These come with a relative weight
of $(-1)^m$. Therefore
\eqn\Wrep{W_n =\sum^{n-1}_{m=0}(-1)^m d_{R_m} e^{-{A\over2}\Lambda(R_m)} .}

Our task is simplified in that we need only the large $N$ limit of \Wrep \
and infinitesimal area. In other words, the sum
\eqn\Winf{W_n(\Delta A) =\sum^{n-1}_{m=0}(-1)^m d_{R_m}
(1-{\Delta A\over2}\Lambda(R_m)) .}
Now $\Lambda(R)$ could, in general, take the 
form \GSY
\eqn\genCas{\Lambda(R)=\sum_{\{k_i\}}a_{\{k_i\}}
C_{\{k_1+2k_2+3k_3+\dots\}}(R)}
where $C_{\{k_1+2k_2+3k_3+\dots\}}(R)$ is a generalised Casimir. 
In the continuum theory this corresponds to \gYM  \ with
\eqn\Phib{\Phi(B)=\sum_{\{k_i\}}a_{\{k_i\}}\prod_i (\tr B^i)^{k_i}.}
For reasons of clarity and computational efficacy we'll henceforth
restrict ourselves to the sufficiently general case
\eqn\Cask{\Lambda(R)=\sum_k {a_k\over N^{k-1}} C_k(R)}
i.e. the higher Casimirs. In the path integral this translates into
\eqn\Phik{\Phi(B)=\sum_k a_k \tr B^k .}
The considerations generalise in a straightforward manner, however.

We can perform the sum \Winf \ with $\Lambda(R_m)$ given by \Cask.
This is undertaken in the appendix. The result is 
\eqn\Un{{1\over N}W_n(\Delta A)=1-{\Delta A\over2}\sum_k a_k 
n{{n+k-2}\choose k-1}}
The S-transform is now readily computed. From \phij 
\eqn\phiqcd{\eqalign{\phi(j)=&\sum_{n=1}^{\infty}(1-{\Delta A\over2}\sum_k a_k 
n{{n+k-2}\choose k-1})j^n \cr
=&{j\over 1-j}(1-{\Delta A\over2}\sum_k a_k {1\over (1-j)^k}
(1+(k-1)j)).}}
According to \Sdef we need the inverse to $\phi(j)$. This is accomplished
easily since we only need it to lowest order in $\Delta A$.
\eqn\chiqcd{\chi_{\Delta A}(z)={z\over 1+z}(1+{\Delta A\over2}
\sum_k a_k(1+kz)(1+z)^{k-2}).}

As argued earlier, $S_A(z)$ for finite area is obtained by
exponentiating $S_{\Delta A}(z)$.
\eqn\Sqcd{\eqalign{S_A(z)= & e^{{A\over2}\varphi(z)} \cr
\varphi(z) \equiv &\sum_k a_k(1+kz)(1+z)^{k-2} \cr
=&\sum_k k(a_k-a_{k+1})(1+z)^{k-1} \cr} .}
We are now in a position to calculate $<\tr [U^n]>$, given the 
explicit form of \Sqcd. From \phij , we conclude that
\eqn\eigen{ <\tr[ U^n]>={1\over 2\pi}\int
\phi(e^{-i\theta})e^{in\theta}d\theta.}
We make the change of variables $\chi(z)=e^{-i\theta}$ exploiting
the fact that $\phi(\chi(z))=z$. Finally after an integration by parts
we reach 
\eqn\Unfin{\eqalign{<\tr[ U^n]>=&{1\over n}\oint 
{dz\over 2\pi i}[\chi_A(z)]^{-n}\cr 
=& {1\over n}\oint {dz\over 2\pi i}(1+{1\over z})^n 
e^{-n{A\over 2}\sum_k a_k(1+kz)(1+z)^{k-2}} .\cr}}

This is the expression that generalises the Laguerre polynomials
that appeared in the usual $QCD_2$ \ref\rossi{P. Rossi,  {\it Ann. Phys.} 
{\bf132}, 463 (1981)}. 
They are, of course, recovered
by restricting ourselves to the case with $a_k=\delta_{k,2}$. We 
also note that the $a_1$ dependence is only through a multiplicative
factor of $e^{-na_1{A\over 2}}$.
The first few averages are 
\eqn\Ufew{\eqalign{<\tr U>=&e^{-{A\over 2}\sum_k a_k};<\tr [U^2]>
=(1-A\sum_k a_k(k-1))e^{-2{A\over 2}\sum_k a_k}\cr
<\tr [U^3]>=&\bigl(1-3{A\over 4}\sum_k a_k(k-1)(k+2)+
3{A^2\over 2}[\sum_k a_k(k-1)]^2\bigr)e^{-3{A\over 2}\sum_k a_k} . \cr}}
The general structure is 
\eqn\Upoly{<\tr[ U^n]>=P_n(A)e^{-n{A\over 2}\sum_k a_k} ,}
where $P_n(A)$ is a polynomial of degree $n-1$ determined by
\eqn\poly{P_n(A)={1 \over (n-1)!}{d^{n-1} \over dx^{n-1}}
[x^n e^{-n{A\over 2}\sum_k k(a_k-a_{k+1})(x^k-1)}]\bigm| _{x=1} .}
As argued in the previous section this is all we need to 
compute the loop average of an arbitrary contour. 

\newsec{The Generalised Hopf Equation and Collective Field Theory}

The Hopf equation arises in the collective field theory 
description in $QCD_2$ \mrd1 . 
It governs the behaviour of the eigenvalue density
of Wilson loops \ref\GM{D.J.Gross, A.Matytsin 
{\it Nucl. Phys.} {\bf B437}, 541 (1995)}. 
In this case we shall find an analogous equation ---
the generalised Hopf equation \us . We shall demonstrate that
\eqn\Fdef{F(\theta, A)\equiv \phi_A(e^{i\theta})
=e^{i\theta}R(e^{i\theta}, A)-1}
obeys the equation 
\eqn\Ghopf{{ \partial F \over \partial A} + {i\over 2}\varphi (F)
{ \partial F \over \partial \theta} = 0,}
with $\varphi (F)$ as in \Sqcd. Here $R(e^{i\theta}, A)$ is the resolvent
\eqn\Resol{R(\zeta, A)=\sum_{n=0}^{\infty}<\tr [U(A)^n]>\zeta^{-(n+1)}.}
We also note that the eigenvalue density of loops $\sigma(\theta, A)$
is given by
\eqn\densty{Re F(\theta, A)=\pi\sigma(\theta, A)-{1 \over 2} .}

If we define $\theta(z, A)$ such that $e^{i\theta(z, A)}=\chi_A(z)$, then
$F(\theta(z, A), A) =z$. This follows from \Fdef \ and the definition of
$\chi_A(z)$ as the inverse to $\phi(j)$. Therefore 
\eqn\hop{{dF(\theta(z, A), A) \over dA}\bigm| _z=
{ \partial F \over \partial A}\bigm| _\theta
+{ \partial F \over \partial \theta}\bigm| _A 
{ \partial \theta(z,A) \over \partial A}\bigm| _z=0.}
Knowing, as we do, $\chi_A(z)$ for generalised $QCD_2$. we see that
\eqn\thetaz{\theta(z, A)={iA\over 2}\varphi(z)- i\ln ({1+z \over z}).}
This, together with \hop \ enables us to arrive at \Ghopf. 

Once again, the hopf equation of $QCD_2$ is regained in the case 
$a_k=\delta_{k,2}$ after a minor shift. Like the Hopf equation,
\Ghopf \ is integrable as well.
In fact, it is easily checked that 
\eqn\Fsoln{F(\theta, A)=F_0(\theta-{iA\over 2}\varphi(F(\theta, A)))}
is the (implicit) solution of this equation. Here $F_0(\theta)=
F(\theta, A=0)$ is the necessary initial condition. The solution is also 
completely determined once we specify, say, $\sigma(\theta,A=0)$
and $\sigma(\theta,A=\infty)$. 

One might imagine obtaining this equation from the analogue of the
loop equations for generalised $QCD_2$. These highly non-linear equations
would have to be derived from a series of Schwinger-Dyson equations for 
loops. This does not seem to be a very easy approach.

Alternatively, one might imagine a derivation of \Ghopf \ from 
a collective field theory description \ref\jevsak{A. Jevicki and B. Sakita,
{\it Nucl. Phys.} {\bf B165}, 511 (1980)}
of generalised $QCD_2$. Logically, one would start from the fermionic
description. The hamiltonian is a rather complicated polynomial in the 
operators ${ \partial^m \over \partial \theta^m}$. One would have to then
bosonise them in their second quantised form and look at the classical
equations of motion.
Here, we have seen it arise much more simply.

In fact, though it was derived here for the plane, we can conclude that the 
equation will be true on the sphere, or more generally, the cylinder.
This follows since these are local evolution equations which are 
sensitive to the global geometry only via the initial/final conditions. 
Thus, on the cylinder we would specify the boundary holonomies 
or eigenvalue densities $\sigma(\theta,A=0)$ and $\sigma(\theta,A=A_0)$
($A_0$ is the area of the cylinder). Then $\sigma(\theta,A)$ would be 
determined from \Ghopf \ with $F(\theta, A)$ a complex function
such that
\eqn\GenF{F(\theta, A)=\pi\sigma(\theta,A)-{1\over 2}-iv(\theta)}
This can be interpreted as  a general kind of fluid flow with density
$\sigma$ and velocity $v$.

But now we can turn the logic around and derive the collective field
theory Hamiltonian whose classical equation of motion is \Ghopf.
It is convenient to redefine variables to 
\eqn\Gdef{G(\theta, A)=1+F(\theta, A)=\pi\sigma(\theta,A)+{1\over 2}
-i\partial _{\theta}\Pi(\theta,A)}
where $\Pi$ is the canonical momentum conjugate to $\sigma$.
Then we have the Poisson bracket
\eqn\Gpb{\{G(\theta, A),G(\theta^{\prime}, A)\}=-i2\pi\partial _{\theta}
\delta(\theta-\theta^{\prime}).}
The generalised Hopf equation now reads as 
\eqn\GGhopf{{ \partial G\over \partial A} + {i\over 2}
\sum_k k(a_k-a_{k+1})G^{k-1}
{ \partial G \over \partial \theta} = 0.}
This is the Hamiltonian equation of motion
\eqn\Gham{\eqalign{ { \partial G\over \partial A}=&\{H,G\}\cr
H=&-{1\over 4\pi}\sum_k {1\over k+1}(a_k-a_{k+1})\int d\theta 
[G^{k+1}(\theta,A)+c.c.] .\cr}}
This is thus the Collective field hamiltonian which can be expressed in terms
of $\sigma(\theta,A)$ and $\Pi(\theta,A)$. Our considerations thus
far have been in euclidean space. The minkowski version reads as
\eqn\Minkham{\eqalign{H_M=&{1\over 2}\sum_k {1\over k+1}(a_k-a_{k+1})
\int d\theta
[({1\over 2}+P_{+})^{k+1}+({1\over 2}-P_{-})^{k+1}]\cr
P_{\pm}=&\partial _{\theta}\Pi(\theta,A) \pm \pi\sigma(\theta,A) .\cr}}
The usual $D=1$ matrix 
model hamiltonian \jevsak 
\eqn\Coll{H={1\over 2\pi}\int d\theta\sigma(\theta)
\{ \partial _{\theta}\Pi(\theta)^2 +{\pi^2 \over 3}\sigma(\theta)^2 \}}
corresponds to taking $a_k=\delta_{k,2}$.

\newsec{The Master Field}

Given that we have computed arbitrary loop averages, we expect to 
have enough information to construct the master field strength. 
To actually do this we will first obtain master loop operators
$\hat{U}(\Delta A)$ for infinitesimal loops. 
These take the form
\eqn\hatu{\eqalign{\hat{U}=&\hat{a}+\sum_{k=0}^{\infty}\omega_{k+1}
\hat a^{\dagger k} \cr <\tr[U^n]>=&
\left\langle \Omega|\hat{U}^n |\Omega\right\rangle .\cr}}
where we need $\omega_{k}$ only to lowest order in $\Delta A$.
Since $U(y)={1\over y}+\sum_{k=0}^{\infty}\omega_{k+1}y^k$
is the inverse of the resolvent, using \Fdef \ and \phij \ we have 
\eqn\Uinv{R({1\over \chi(z)})=(1+z)\chi(z) \Rightarrow
U((z+1)\chi(z)) ={1\over \chi(z)}.}
Therefore \Sqcd \ implies
\eqn\Uexp{\eqalign{U(y)=&{1\over y}(1+z(y));\ \  y=ze^{{A\over 2}\varphi(z)}\cr
\Rightarrow &\sum_{k=0}^{\infty}\omega_{k+1}y^{k+1}=z(y)}}
This is easily solved to lowest order in $A$ to obtain
\eqn\zy{z(y)=y(1-{\Delta A\over 2}\varphi(y)).}
So the master loop operator $\hat{U}(\Delta A)$ is 
\eqn\hatufin{\hat{U}(\Delta A)=\hat{a}+1-{\Delta A\over 2}
\varphi(\hat{a}^{\dagger}).}

Let us now construct the operator $\hat{H}$ such that $e^{i\hat{H}}=\hat{U}$.
This is best done as follows. 
\eqn\Uh{\hat{U}^n=e^{in\hat{H}}=1+in\hat{H}-{n^2\over 2}\hat{H}^2+ \ldots .}
Therfore $\hat{H}$ is identified as the linear term in $n$ in 
$\hat{U}^n$. Again matters are simplified in having to keep only
terms of $O(\Delta A)$. Then from \hatufin
\eqn\Uexp{\eqalign{\hat{U}^n=&(\hat{a}+1-{\Delta A\over 2}
\varphi(\hat{a}^{\dagger}))^n \cr
=&(\hat{a}+1)^n-{\Delta A\over 2}\sum_{m=0}^{n-1}(\hat{a}+1)^k
\varphi(\hat{a}^{\dagger})(\hat{a}+1)^{n-1-k} \cr}}

First look at the term of order $\Delta A$.
Terms in this expansion of the form $\hat{a}^{\dagger l}\hat{a}^m$
do not contribute to $\hat{H}^n$ in leading order in $\Delta A$.
We group the other terms where $\hat{a}^m$ multiplies on the left
and look for the contribution linear in $n$.
This gives us the term
\eqn\term{\sum_{r=0}{(-1)^{r}\over (r+1)}\hat{a}^r\varphi(\hat{a}^{\dagger}).}
In this expression we can drop the terms purely of the form $\hat{a}^m$.
Since there are already such terms of $O(1)$ and the $O(\Delta A)$
only contribute to higher order. The $O(1)$ terms can be easily read from the 
$A=0$ limit when $\hat{U}=\hat{a}+1$. Putting it all together
\eqn\hath{\eqalign{\hat{H}=&-i(log(1+\hat{a})\{1+{\Delta A\over 2}
\hat{a}^{\dagger}\varphi(\hat{a}^{\dagger}\}\cr
=&\sum_{r=1}^{\infty}{(-1)^{r}\over r}\hat{a}^r
+{\Delta A\over 2}\sum_n c_n \hat{a}^{\dagger n}\cr
c_n=&\sum_k a_k \sum_{r=0}^{k-n-1}{(-1)^{r}\over (r+1)}
(n+r+1){{k-1}\choose n+r} \cr}}
We notice that $c_0=0$.

Matters are simplified if we choose axial gauge $A_1=0$. Then
the holonomy for an infinitesimal loop becomes
\eqn\hol{U=P\exp (i\oint_{C} A_{\mu}dx^{\mu})=\exp (i\partial_1A_0 \Delta A).}
We can then identify the large $N$ limit of the two matrices:
\eqn\field{\hat{F}\Delta A=\partial_1A_0 \Delta A=\hat{H}}
with $\hat{H}$ as in \hath. Since $c_0=0$, $\hat{F}$ has a vanishing
1-point function which is expected. The presence of the $\Delta A$
is a consequence of discretisation of the action. This ensures that
correlation functions of the field strength $F$ do not have delta 
functions in them. In this gauge the master field $\hat{A_1}=0$
and $\hat{A_0}$ is determined by \field and \hath .

\newsec{Discussions and Conclusions}

In this work we have applied the formalism of the master field to 
explicitly compute the large $N$ limit of many quantities in generalised
$QCD_2$. This was meant to illustrate the utility of the framework of
non-commutative probability theory in actual computations. As we have 
seen, it is not easy to see how to obtain the collective field theory 
description or the expressions for loop
averages and gauge fields  with the usual means available to us. 
It is worthwhile to make some remarks regarding them here:
The collective field equations of motion  of Section 4 are related
to the non-commutative probability distributions in the same way as
the heat equation is related to the Gaussian distribution. This gives
us an inkling as to the mathematical relation of the collective field
description and the underlying matrix model. It might help to
understand the relation better so as to arrive at tractable
collective field descriptions of realistic theories. On a technical 
level, we note the simplification gained in going to an operator 
construction of the master field. In Section 5 we were able to go from
loop averages to gauge fields, simply by performing standard operator
manipulations like taking the logarithm. Analogous operations in a 
large $N$ matrix description do not appear easy.

We have, of course, in this paper,
exploited a very special property of this system, namely that of being 
multiplicatively free. This would not straightforwardly generalise to 
higher dimensions. Nevertheless, the lesson that is perhaps to be drawn 
is that techniques can be developed in the operator framework for the 
master field which can take us beyond conventional approaches.  

\bigskip
{\bf Acknowledgements:} I would like to thank David Gross for suggesting
the applicability of the master field framework in this context, as 
well as for useful discussions. I would also like to acknowledge 
Ori Ganor for helpful conversation.   

\bigskip
{\bf Appendix}

Here we undertake to perform the sum 
\eqn\apsum{W_n(\Delta A) ={1\over N}\sum^{n-1}_{m=0}(-1)^m d_{R_m}
(1-{\Delta A\over2}\Lambda(R_m))}
to leading order in large $N$, over representations $R_m$ of \YT.
The dimension of the representation is given by 
\eqn\dim{d_{R_m}=\prod_{i>j}(1+{h_j-h_i \over i-j})=
{1\over n}{(N+n-m-1)!\over (N-m-1)!m!(n-m-1)!} .}
$\Lambda(R_m)$ is given by \Cask , and the higher Casimirs are in general 
given by 
\eqn\HighC{\eqalign{C_k(R)=&\sum_{i=1}^N l_i^k \gamma _i ,\cr
l_i=& h_i+N-1; \ \ \ \gamma _i=\prod_{i \neq j}(1-{1\over l_i-l_j}). \cr}}
In our case \YT, this can be computed to be
\eqn\casm{\eqalign{C_k(R_m)=&l_1^k \gamma _1 +l_{m+1}^k \gamma _{m+1} \cr=&
{n \over n-1}[(N+n-m-1)^{k-1}(n-m-1)+(N-m)^{k-1}m] \cr
=&{n \over n-1}[(N+n-m-1)^k-(N-m)^k -N (k\rightarrow k-1)] .\cr }}
Using \dim \ it is easy to show that the $O(1)$ term in \apsum \ is indeed $1$.
The non-trivial part is the $O(\Delta A)$ term which reads
\eqn\ntsum{\eqalign{\sum_{k>0}{a_k \over N^k}{n\over n-1}&\sum^{n-1}_{m=0}
(-1)^m{N+n-m-1 \choose n}{n-1 \choose m}\cr
&\times [(N+n-m-1)^k-(N-m)^k -N (k\rightarrow k-1)]. \cr }}
For fixed $k$ consider the $(N-m)^k$ term. Expanding in powers of $N$, 
the sum can be expressed as the coefficient of $y^{N-1}$ in
\eqn\asum{\eqalign{&{n\over n-1}\sum_{r=0}^k(-1)^r{k\choose r}
N^{k-r}\sum^{n-1}_{m=0}(-1)^m{n-1 \choose m}(1+y)^{N+n-m-1}
(z{d\over dz})^r z^m |_{z=y, y^{N-1}}\cr
&=(-1)^{n}{n\over n-1}\sum_{r=0}^k(-1)^r{k\choose r}
N^{k-r}(1+y)^{N}(z{d\over dz})^r (z-y-1)^{n-1}|_{z=y, y^{N-1}} .\cr}}

Next we write $(z{d\over dz})^r=\sum_{l=1}^r C_l^r z^l{d\over dz}^l$,
where $C_l^r$ are coefficients determined by the recursion relation
\eqn\crl{C_l^r=lC_l^{r-1}+C_{l-1}^{r-1}; \ \ C_r^r=1 .}
Thus \asum \ reads as
\eqn\bsum{{n\over n-1}\sum_{r=0}^k(-1)^r{k\choose r}
N^{k-r}(1+y)^{N}\sum_{l=1}^r C_l^r (-1)^ly^l(n-1)(n-2)\ldots(n-l)|_{y^{N-1}}.}
Since to leading order in $N$ we only need the term that goes as 
$N^k$, it suffices to look for the coefficient of $N^r$ in 
\eqn\Nrterm{(1+y)^{N}y^l|_{y^{N-1}} ={N \choose l+1}
={N^{l+1}\over (l+1)!}+{N^l\over 2(l-1)!}+O(N^{l-1}).}
Thus only the $l=r$ and $l=r-1$ terms contribute in the sum over $l$ in 
\bsum. Therefore, using $C_r^r=1$ and $C_{r-1}^r={r(r-1)\over 2}$, 
the leading contribution to \bsum \ simplifies into  
\eqn\csum{-{n\over 2}N^k\sum_{r=0}^k{k\choose r}{n-1\choose r-1}
=-{n\over 2}N^k{n+k-1\choose n}.}
It can be checked that the $(N+n-m-1)^k$ term in \ntsum \ gives a 
contribution of opposite sign but same magnitude as \csum. Therefore the 
full contribution from \ntsum \  reads as 
\eqn\finsum{\sum_{k>0}a_kn[{n+k-1\choose n}-{n+k-2\choose n}]
=\sum_{k>0}a_kn{n+k-2\choose n-1} .}
This is the result used in \Un . 

\listrefs

\end